# A Causal-Guided Multimodal Large Language Model for Generalized Power System Time-Series Data Analytics

Zhenghao Zhou, *Student Member*, *IEEE*, Yiyan Li, *Member*, *IEEE*, Xinjie Yu, Runlong Liu, Zelin Guo, Zheng Yan, *Senior Member*, *IEEE*, Mo-Yuen Chow, *Fellow*, *IEEE*, Yuqi Yang and Yang Xu

***Abstract*—Power system time series analytics is critical in understanding the system operation conditions and predicting the future trends. Despite the wide adoption of Artificial Intelligence (AI) tools, many AI-based time series analytical models suffer from task-specificity (i.e. one model for one task) and structural rigidity (i.e. the input-output format is fixed), leading to limited model performances and resource wastes. In this paper, we propose a Causal-Guided Multimodal Large Language Model (CM-LLM) that can solve heterogeneous power system time-series analysis tasks. First, we introduce a physics-statistics combined causal discovery mechanism to capture the causal relationship, which is represented by graph, among power system variables. Second, we propose a multimodal data preprocessing framework that can encode and fuse text, graph and time series to enhance the model performance. Last, we formulate a generic "mask-and-reconstruct" paradigm and design a dynamic input-output padding mechanism to enable CM-LLM adaptive to heterogeneous time-series analysis tasks with varying sample lengths. Simulation results based on open-source LLM Qwen and real-world dataset demonstrate that, after simple fine-tuning, the proposed CM-LLM can achieve satisfying accuracy and efficiency on three heterogeneous time-series analytics tasks: missing data imputation, forecasting and super resolution.**

## I. Introduction

DRIVEN by global energy transition and the strategic pursuit of carbon neutrality, modern power systems are undergoing a profound transformation. The large-scale integration of renewable energy sources and flexible loads has not only increased the complexity of grid dynamics, but also generated massive volumes of high-dimensional time-series data enabled by advanced metering infrastructure [1].

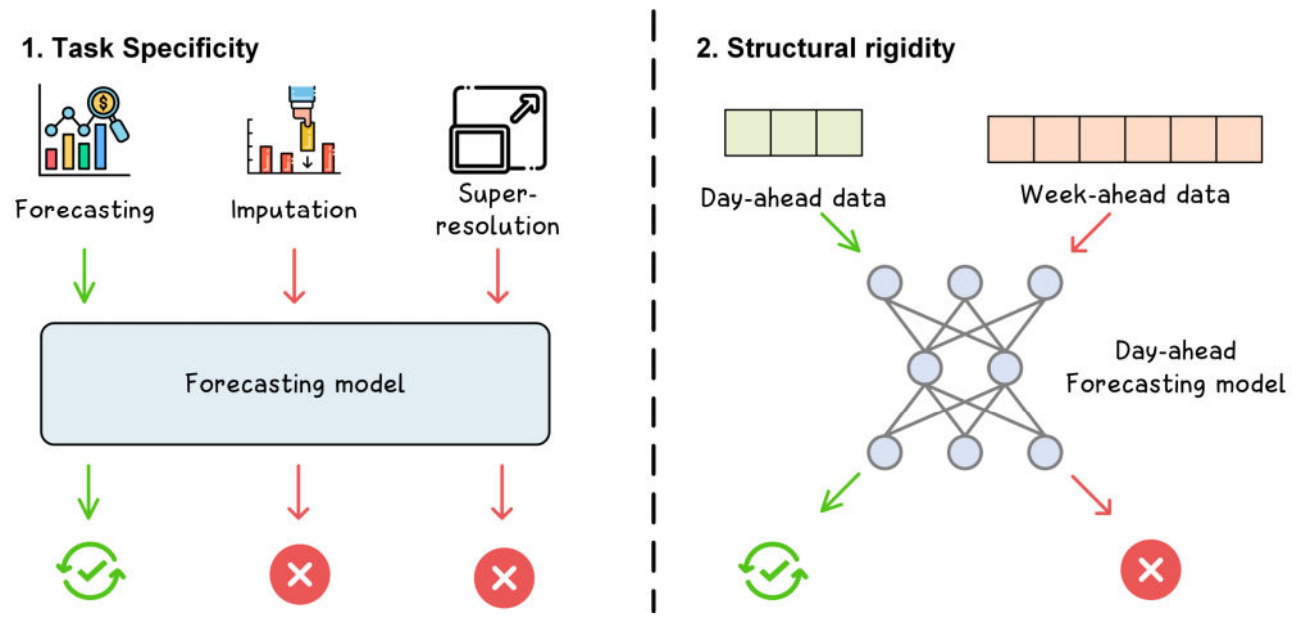

Fig. 1. Illustration about task specificity and structural rigidity problems for traditional AI models.

As such, extracting operational insights from the large scale but often noised and incomplete datasets have become a key challenge in modern power systems. These tasks can be collectively referred to as time-series data analysis in power systems. For example, data imputation aims to ensure data integrity by recovering missing or corrupted values; short-term forecasting provides decision support for day-ahead scheduling by predicting future system states; super-resolution enhances temporal resolution to support fine-grained modeling and control.

To handle the time series data analysis problem, Artificial Intelligence (AI), particularly deep learning models, has gained increasing attention in power systems, demonstrating significant achievements in enhancing system reliability and efficiency [2]-[12]. However, as illustrated in Fig. 1, most existing methods suffer from the following two limitations, which impair their real-world implementation:

- **Task specificity:** Most AI-based models are designed for a specific task (e.g., forecasting) with tailored architectures and optimization objectives, making them difficult to be generalized to other tasks.
- **Structural rigidity:** Models require fixed input-output dimensions, struggling in solving flexible problems with varying input-output formats (such as varying forecasting horizons, missing data lengths, etc.).

In recent years, the emergence of foundation models, especially large language models (LLMs), offers a promising new paradigm for solving this dilemma. Based on a single pre-trained backbone, LLMs demonstrate unprecedented generalization capabilities that can support a wide range of downstream tasks, paving the way for realizing a truly generalized time series analysis framework in power systems [3]-[27]. Currently, there are two technical mainstreams of leveraging LLMs for domain-specific general-purpose modeling:

- **Domain-specific pretraining**: This approach uses power system domain knowledge to train a LLM from scratch. Such models are highly customized to the power systems and may generate very professional

This work was supported by National Natural Science Foundation of China under Grant 52307121, supported by Shanghai Sailing Program under Grant 23YF1419000, and also supported by the Open Research Fund of National Engineering Research Center of Water Resources Efficient Utilization and Engineering Safety. (Corresponding author: Yiyan Li.)

Zhenghao Zhou, Yiyan Li, Xinjie Yu, Runlong Liu, Zelin Guo are with the College of Smart Energy, Shanghai Non-Carbon Energy Conversion and Utilization Institute, and Key Laboratory of Control of Power Transmission and Conversion, Ministry of Education, Shanghai Jiao Tong University, Shanghai, 200240, China. (e-mail: zhenghao.zhou@sjtu.edu.cn, yiyan.li@sjtu.edu.cn, yuxinjie@sjtu.edu.cn, runlong_liu@sjtu.edu.cn, gzl1996@sjtu.edu.cn).

Zheng Yan is with the Key Laboratory of Control of Power Transmission and Conversion, Ministry of Education, and the Shanghai Non-Carbon Energy Conversion and Utilization Institute, Shanghai Jiao Tong University, Shanghai 200240, China. (e-mail: yanz@situ.edu.cn)

Mo-Yuen Chow is with the Global College, Shanghai Jiao Tong University, Shanghai, 200240, China. (email: moyuen.chow@sjtu.edu.cn)

Yuqi Yang, Yang Xu are with China Yangtze Power Co., Ltd. (CYPC), Wuhan, 430000 , China. (email: yang_yuqi1@ctg.com.cn, xu_yang@ctg.com.cn)

results. However, pretraining a power system LLM requires huge amounts of domain-specific data samples and extensive computation costs, limiting the models' parameter scale and overall capability compared with leading commercial or open-source LLMs. Along this line, Wang et al. propose a novel unified pre-trained model for smart meter data, which can learn general load patterns through pre-training [28]. However, different types of tasks still require task-specific output layers and fine-tuning strategies.

- **Adaptation of pretrained LLMs**: This approach is based upon the state-of-the-art general-purpose LLMs that have been pretrained at great cost, such as Qwen, Llama, and ChatGPT. These models have powerful capabilities in sequence modeling, instruction following, and multimodal processing, which can be adapted to time-series analysis tasks through techniques such as supervised fine-tuning. This approach can significantly lower the development costs of domain-specific LLMs, and has yield promising results in recent studies [29].

In this paper, by strategically adapting the pretrained LLMs, we develop a Causal-guided Multimodal LLM (CM-LLM), for generalized power system time series data analysis purposes. The time series analysis tasks are formulated as a generic "mask-and-reconstruct" perspective. Multimodal data including text, graph and time series are combined as the model inputs to enhance the performances. Specifically, the text modality consists of instruction prompts that can guide the model's behavior; the graph modality represents causal relationships among variables; and the time series data are real-world multivariate observations to be analyzed. To effectively process these heterogeneous data types, the proposed model employs specially-designed embedding layers for each modality, followed by a fusion mechanism that integrates the multi-source embeddings into a unified representation. Meanwhile, to address the structural rigidity issue, the proposed model supports variable-length input and output sequences through a customized dynamic padding mechanism, enabling efficient batch processing of data with diverse lengths. The entire process requires only a single fine-tuning stage, allowing a single model to effectively handle multiple tasks with minimal adaptation costs. As an example, three different time series analysis tasks: data imputation, forecasting and super-resolution are selected to demonstrate the generic analytical capabilities of the proposed method.

Our main contributions are summarized as follows:

1. A physics-statistics combined causal discovery mechanism: We propose a novel causal discovery mechanism that can effectively capture the causal relationship among power system time series variables. This mechanism leverages the physical connections among variables as the backbone of the causal relationship, which is then optimized by the Peter-Clark (PC) algorithm based on field measurements. As such, this mechanism has good interpretability and reliability in causal structure learning, providing an important modality input to the proposed CM-LLM model.

2. A unified multimodal data preprocessing architecture: We design a data preprocessing architecture that can unify multimodal data inputs, including textual instructions, graph structures (representing causal relationships) and multivariate time series. In this architecture, each modality is encoded into vectors respectively, followed by specially-designed data fusion mechanism. As such, this data preprocessing architecture can enable the proposed CM-LLM model to gain a deep understanding about complex power system patterns and tasks.

3. A generic task representation paradigm: We propose a generic "mask-and-reconstruct" paradigm to uniformly represent different time series analysis tasks, including but not limited to missing data imputation, forecasting, and super-resolution. Such a task representation paradigm can enable a single LLM to solve different time series analysis tasks, fully unleash the generalization capabilities of LLMs so that the task specificity issue can be solved.

4. A dynamic input-output padding mechanism: We propose a dynamic input-output padding mechanism, including dynamic batch-wise padding and sequence-aware loss computation, to handle the varying input-output sequence lengths of different time series analysis tasks. As such, the structural rigidity issue can be solved without sacrificing model accuracy.

The remainder of this paper are organized as follows: Section II reviews the related works; Section III presents the methodology; Section IV demonstrates the case study results; and Section V provides the conclusions.

## II. Related Work

### A. Power System Time-Series Data Analysis

Time-series data analysis enhances operators' understanding of grid situation and serves as the foundation for decision-making process. Currently, most time-series analysis models are designed for specific tasks and are trained in end-to-end fashion on small-scale datasets [3]. For example, to capture temporal dynamics, recurrent neural networks, such as Long Short-Term Memory (LSTM)[4] and gated recurrent units [5], have been widely adopted in forecasting tasks. Generative models, particularly those based on generative adversarial networks and diffusion processes, have also been implemented in time-series analysis tasks such as forecasting [6], data imputation [7] and super-resolution [8] due to strong distribution-learning ability. In addition, temporal convolutional networks [9] and Transformer-based architectures [10] have demonstrated superiority in modeling long-term temporal dependencies. Considering that power system is essentially a multivariate system, researchers have introduced auxiliary information such as graph structure [11] and cross-variable correlations [12] to enhance the time-series analysis performances. While these approaches have achieved promising results on individual tasks, they often lack generalization and adaptation capabilities to heterogeneous tasks (i.e., task specificity issue) and are constrained by fixed input-output configurations (i.e., structural rigidity issue).

### B. LLM Applications in Power Systems

LLMs have recently gained increasing attention in power systems due to their versatile modeling and generalization capabilities [13]. Based on the model pretraining strategy, Tu et al. establish a hierarchical-temporal model pretrained on electricity time series data, achieving superior performance across downstream tasks [14]. Zhu et al. introduce a physics-informed graph transformer that leverages grid topologies and dynamic mixture-of-expert layers [15]. Hu et al. propose BERT-PIN, a framework that leverages natural language processing techniques for power load restoration by encoding time-series data as linguistic tokens in a learned "language" of consumption patterns [16]. Other researchers adapt existing pre-trained LLMs through techniques such as fine-tuning [17], in-context learning [18], and retrieval-augmented generation [19]. In addition, some researchers have explored the use of LLMs as intelligent assistants to support decision-making processes. For instance, LLMs have been employed to recommend predictive models [23], generate code [24], [25], produce reinforcement learning strategies [26], or provide suggestions for optimal operational adjustments [27].

### C. Causal Analysis in Power Systems

Causal relationships represent intrinsic and invariant dependencies among variables that remain stable across different conditions [30]. Because power system is a typical multivariate physical system, uncovering such causal structures can provide a more reliable and interpretable basis for data analysis [31], [32], [33]. For example, Zhao et al. propose CEDAN, a causality-enabled domain adaptation network that extracts invariant causal features from pre-trained source PV data, enabling unsupervised zero-label forecasting for new photovoltaic sites [34]. Jiang et al. show that causal meta-learners, pre-trained on debiased observational data via propensity trimming, can accurately estimate HVAC impacts on building thermal dynamics [35].

## III. Methodology

In this section, we select three typical time-series analysis tasks: data imputation, forecasting and super-resolution, as examples to demonstrate how the proposed method can unify them into a common representation. Next, we introduce the proposed CM-LLM framework, which enables generalized time series analysis for power systems through a single training process.

### A. Problem Formulation

As illustrated in Table I and Fig. 2, despite the differences among the tasks of data imputation, forecasting and super-resolution, their core objective is fundamentally the same: to infer and reconstruct unknown information by learning the patterns from the observed time series.

If we treat the unknown regions as masked segments, the three different tasks can be uniformly formulated as a "mask-and-reconstruction" process. In the case of imputation, the mask corresponds to missing segments in the time series, and the goal is to reconstruct these missing parts using contextual information. For forecasting, the mask covers all future time steps, and the model infers future trends based on historical observations. In super-resolution, the mask represents the implicit high-resolution points to be recovered between low-resolution measurements.

Given that power systems are typically multivariate in nature, we can thus obtain a unified problem formulation for these three tasks as follows:

Denote a complete multi-variable time series data matrix $\mathbf{X}$, including system and explanatory variables as

$$\mathbf{X}=\begin{bmatrix} x_1^1 & x_2^1 & \cdots & x_L^1 \\ x_1^2 & x_2^2 & \cdots & x_L^2 \\ \vdots & \vdots & \ddots & \vdots \\ x_1^E & x_2^E & \cdots & x_L^E \end{bmatrix} \quad (1)$$

where $E$ is the number of variables and $L$ is the length of the time series. We define a mask matrix $\mathbf{M}$, which consists of binary values, 0 and 1. The size of $\mathbf{M}$ is similar to $\mathbf{X}$.

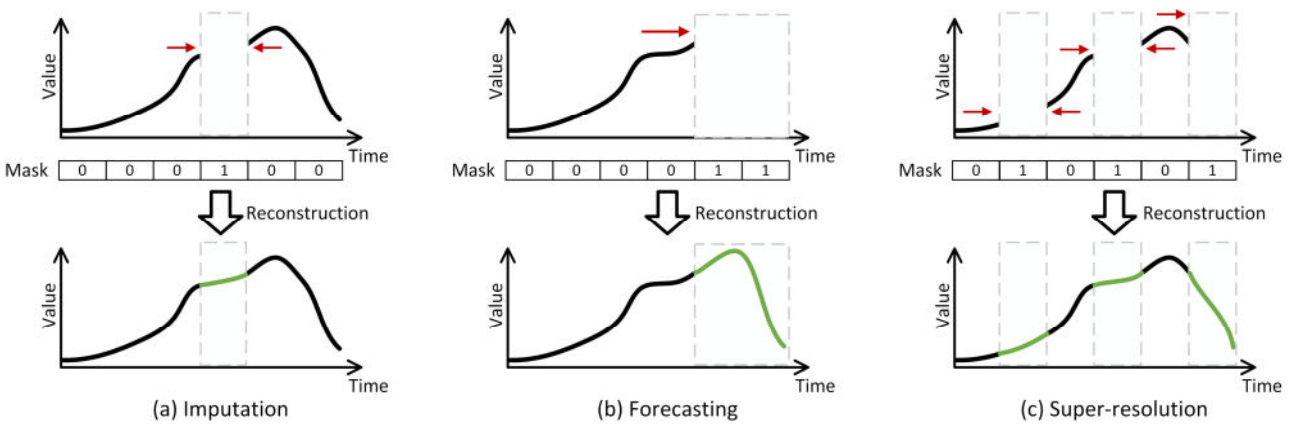

Fig. 2. The "mask-and-reconstruction" generic representation paradigm.

TABLE I
SUMMARY OF THREE TASKS

| Task | Imputation | Forecasting | Super-resolution |
|---|---|---|---|
| Objective | Restore data integrity | Infer future states | Increase data granularity |
| Unknown position | Missing segments within a known time range | Future segments beyond the known time range | Samples between low-resolution points |
| Data context | Bi-directional | Uni-directional | Bi-directional |

In the mask matrix, we designate segments to be marked as 1. Therefore, input data $\tilde{\mathbf{X}}$ can be viewed as a masked version of the underlying complete multivariate time series, where certain entries are missing or unobserved:

$$\tilde{\mathbf{X}}=(\mathbf{1}-\mathbf{M})\circ\mathbf{X} \quad (2)$$

Thus, problem can be described as:

$$\hat{\mathbf{X}}=f_\theta(\tilde{\mathbf{X}}), \text{with } Loss(\hat{\mathbf{X}},\mathbf{X}) \quad (3)$$

where $f_\theta$ is the mapping function. The goal is to find a function, $f_\theta$, to reconstruct the complete multi-variable time series data.

### B. CM-LLM Model Architecture

As illustrated in Fig. 3, the proposed CM-LLM model comprises three fundamental stages: multimodal data preprocessing, multimodal information fusion, and CM-LLM inference. Through the Supervised Fine-Tuning (SFT), the model can understand the time series patterns and are able to transform the multi-source heterogeneous inputs to diverse outputs.

In the multimodal data preprocessing stage, we collect input data from multiple modalities and embed them into an LLM-oriented pattern. This is because LLM is originally designed for Natural Language Processing (NLP) tasks, which expects inputs in the form of tokenized word embeddings. Therefore, during the input adaptation process, we need to generate embeddings for each modality accordingly. First, textual data are derived from natural language descriptions of tasks and scenario instructions, which are then converted into prompt embeddings. Second, raw power system time-series data are processed through a masking mechanism and transformed into unified temporal embeddings. Third, prior causal graphs are refined using the PC algorithm and subsequently encoded into causal embeddings via a Graph Neural Network (GNN).

The second stage of the framework is multi-modal information fusion, wherein embeddings derived from diverse data modalities are systematically aligned and effectively integrated. First, temporal embeddings and causal embeddings are combined through an element-wise summation operation, aiming to enrich the temporal feature representation with encoded variable interactions. The resulting embedding vector is then concatenated with the prompt embeddings, which are generated from task-specific textual instructions, to construct a comprehensive sequence representation. Such integrated representation not only encapsulates rich contextual and semantic information but also incorporates structured guidance for the target task, thereby facilitating the model's reasoning and decision-making capabilities within the given application scenario.

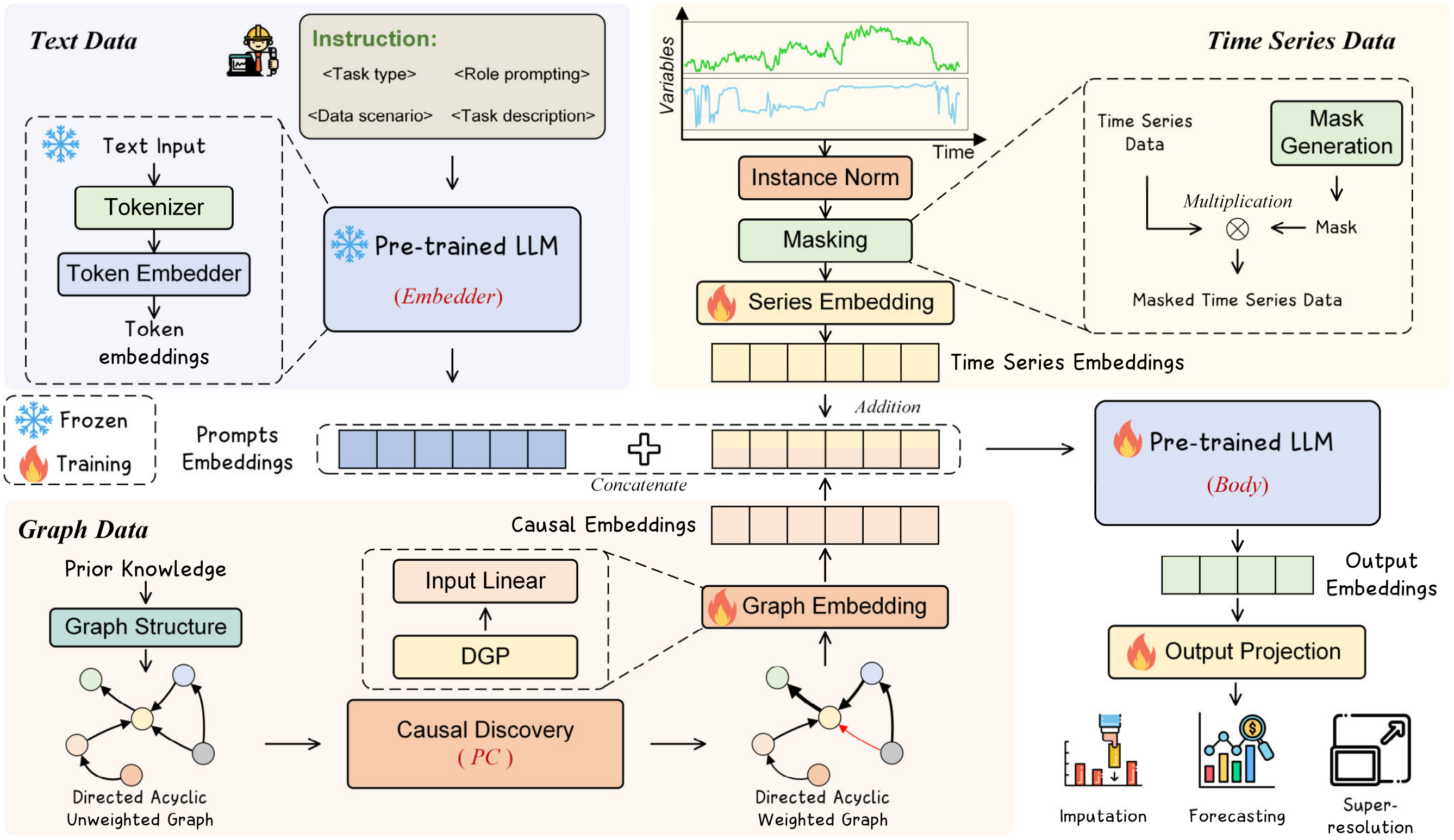

Fig. 3. The model framework of CM-LLM.

In the final CM-LLM inference stage, the fused sequence is input into a pre-trained large language model (LLM), which functions as the core reasoning module for capturing complex nonlinear dependencies and long-term temporal dynamics. The hidden state sequence generated by the LLM is then extracted. Using the known length of the instruction prompt, a slicing operation isolates the segment corresponding to the time series. Subsequently, this segment is mapped to the target output dimension through a linear projection layer.

In the following sections, we will introduce the above three main components in detail.

### C. Multimodal Data Preprocessing

#### 1) Textual Modality: Instruction Prompts

Prompt engineering is a methodology in natural language processing that focuses on designing effective prompts for language models. It aims to guide models to generate desired outputs by formulating specific instructions or queries. Well-crafted prompts can significantly enhance model performance across a variety of tasks, such as data restoration. As illustrated in Fig. 3, the instruction can be regarded as a structured prompt in the context of supervised fine-tuning, which incorporates four types of information to guide the model's behavior. The task type specifies the nature of the operation, while the data scenario provides contextual information regarding the physical background, data modality, and operational conditions. Role prompting helps frame the model's expected function, such as acting as a predictor or causal analyzer, and the task description offers a detailed natural language specification of the problem to be addressed. Together, these elements enable the model to interpret and execute complex temporal analysis tasks in a semantically coherent and context-aware manner.

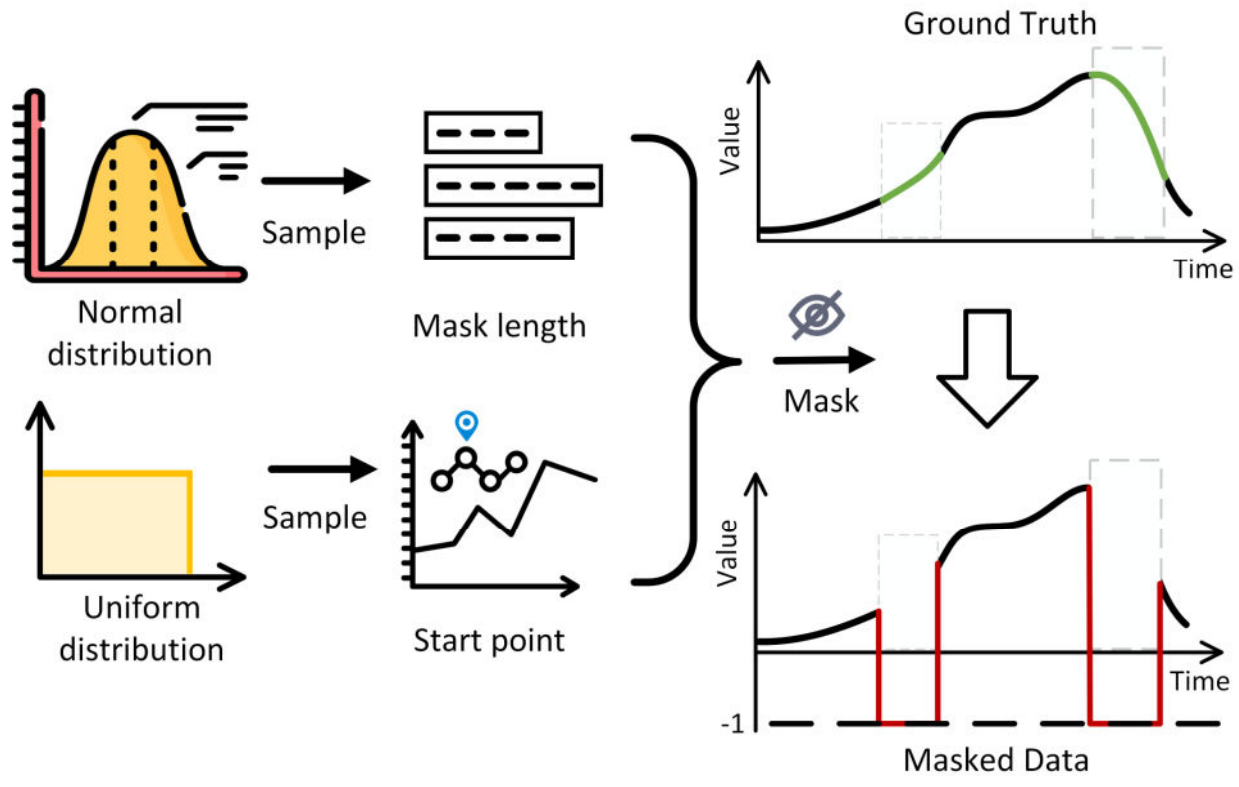

Fig. 4. The masking method for data imputation.

#### 2) Temporal Modality: Multivariate Time Series

Multivariate time series data constitutes the main part of the SFT dataset. The processing procedure is as follows:

First, each temporal variable is individually normalized using instance normalization, transforming the data to lie within the interval [0,1] to obtain the standardized dataset $\mathbf{X}$.

Second, considering that mask positions are different in tasks, 3 strategies for generating mask matrix $\mathbf{M}$ are designed for different tasks. Among the tasks considered, imputation involves the most complex masking strategy. Specifically, as shown in Fig. 4, we randomly select continuous time segments in $\mathbf{M}$ and mark them as missing (i.e., assigned a mask value of 1), where the length of each masked segment is sampled from a normal distribution. The parameters of the distribution are tunable and can be adapted according to the specific characteristics of the dataset in use. The process can be formulated:

$$\ell = \max\left(0, \left\lfloor \mathcal{N}(\mu, \sigma^2) \right\rfloor\right);\ s \sim \mathcal{U}\left(1, L - \ell_i + 1\right) \tag{4}$$

$$M[s : s + \ell_i - 1] \leftarrow 1 \tag{5}$$

where $\ell$ is the length of the missing data segment in imputation task, $\mathcal{N}(\mu, \sigma^2)$ is the normal distribution with a mean of $\mu$ and a variance of $\sigma^2$, $s$ is the started point of missing data segment, $\mathcal{U}$ is a uniform distribution and $L$ is the length of total segment.

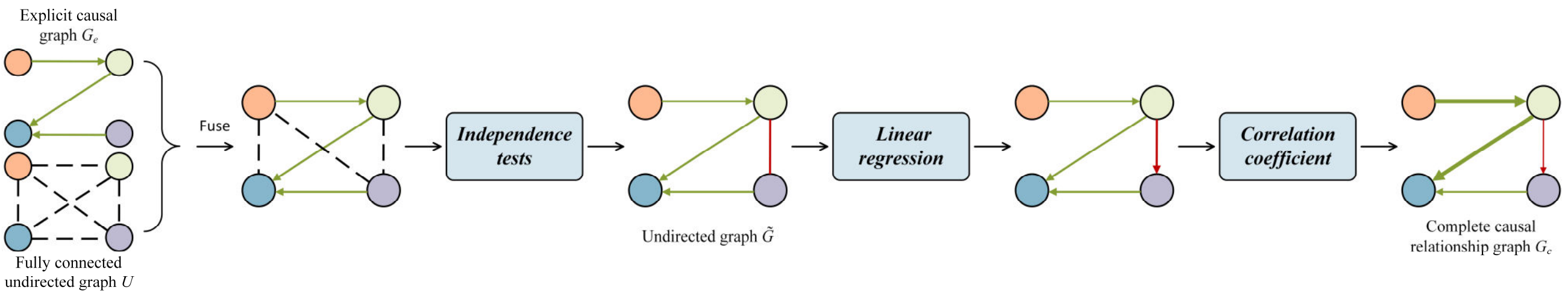

Fig. 5. The process of the physics-statistics combined causal discovery mechanism.

Unlike imputation where missing segments can occur anywhere in the time series, the endpoint in a forecasting task is fixed, and the forecasting window is also fixed. In this case, we only need to determine the length of the future window to be predicted. The values within the prediction window are masked out by the mask matrix $\mathbf{M}$. For super-resolution, the generation of the mask matrix $\mathbf{M}$ is straightforward: all positions in the mask matrix corresponding to the intermediate points between low-resolution observations are set to 1. The effects of mask matrix are two-fold: (1) It is used to mask true values in the input data, thereby facilitating the construction of SFT training pairs; (2) It serves to identify the masked positions and facilitates the calculation of the loss function by focusing on the relevant regions of the input.

Third, based on the mask matrix, all elements in the dataset $\mathbf{X}$ that correspond to positions marked as 1 in the mask are replaced with -1. It is worth noting that, since the data in $\mathbf{X}$ have already been normalized to the range [0, 1], the value -1 serves as a distinct and easily identifiable indicator of masked positions.

$$\tilde{\mathbf{X}} = (\mathbf{1} - \mathbf{M}) \circ \mathbf{X} + (-1) \cdot \mathbf{M} \tag{6}$$

*3) Graph Modality: Causal Relationship*

Prior knowledge of multivariate physical systems, derived from governing physical laws, dictates explicit causal relationships among variables. We encode this knowledge as a Directed Acyclic Unweighted Graph (DAUG), where nodes represent system variables and directed edges denote these causal dependencies.

Rooted in stable physical constraints, the DAUG serves as a foundational causal hypothesis, providing a structured, computationally tractable framework derived from domain expertise. To refine this initial structure and quantify causal strengths, we employ a physics-statistics mechanism. This approach utilizes the PC algorithm, which applies conditional independence testing to observational data to infer a more comprehensive causal graph.

The core procedure of the mechanism proceeds as shown in Fig. 5 and Algorithm 1:

First, a fully connected undirected graph $U$ is initialized, where every pair of variables is connected by an undirected edge. Then, for each pair of variables, the Fisher Z conditional independence tests are conducted under varying conditioning sets. If a pair of variables is found to be conditionally independent given a certain set, the corresponding edge is removed. Notably, explicitly defined causal edges $G_e$ derived from prior knowledge are preserved without undergoing such tests. For the remaining edges, the algorithm further evaluates the predictive power between variable pairs using linear regression models. Specifically,

**Algorithm 1:** Physics-statistics combined causal discovery

**Input:** Explicit causal graph $G_e$, supervised fine-tuning dataset $D_s$.
**Output**: Complete causal relationship graph $G_c$.

```
Procedure:
  Initialize fully connected undirected graph U.
  for each edge (u, v) ∈ U.edges() do:
    if (u, v) ∈ Ge.edges() :
      break
    else
      Perform Fisher Z conditional independence test
      if independent:
        Remove edge (u, v)
      else
        Remain edge (u, v)
      end
    end
  end
  return undirected graph G̃
  for each edge (u, v) ∈ G̃.edges() do:
    Fit u→v and v→u via linear regression.
    Set direction by higher R².
  end
  return directed graph Ĝ
  for each edge (u→v) ∈ Ĝ.edges():
    Compute Pearsonr (u, v | Z) with parents Z of v.
    Assign as edge weight.
  end
  return Gc
```

variable $u$ is used to predict $v$, yielding the coefficient of determination $R^2_{uv}$, and vice versa for $R^2_{vu}$. By comparing these two values, the direction of the edge is assigned from the variable with higher predictive power to the one with lower predictive power.

Throughout this process, all edges representing explicitly defined causal relationships based on prior knowledge are retained in the final graph without participating in the conditional independence tests. This ensures effective utilization of prior knowledge while also improving computational efficiency. The resulting output is a Directed Acyclic Graph (DAG) that incorporates both prior-guided and data-validated causal relationships. To further quantify the significance of each causal edge, we compute weights for every directed edge ($u \rightarrow v$) in the resulting DAG. Specifically, for the edge ($u \rightarrow v$), we construct a conditioning set $\mathbf{Z}$ consisting of other parent nodes of $v$, and calculate the absolute value of the partial correlation coefficient. The absolute value of this partial correlation serves as the weight of the edge, reflecting the strength of the causal influence of $u$ on $v$.

*4) Fused Representation: Multimodal Embedding*

The three heterogeneous data modalities are encoded via different embedding approaches, and the corresponding representations are integrated into a cohesive input for the LLM. The whole instruction text $I_{txt}$ is embedded by the tokenizer and token embedder of the LLM to prompt embeddings. Through fully connected layers, the time series data $\tilde{\mathbf{X}}$ is embedded into temporal embeddings. The embedding process can be formulated:

$$E_{txt} = f_{embed}(I_{txt}) \in \mathbb{R}^{B \times L_{txt} \times D_{hidden}} \quad (7)$$

$$E_{ts} = f_{ts}(\tilde{\mathbf{X}}) \in \mathbb{R}^{B \times L_{ts} \times D_{hidden}} \quad (8)$$

where $f_{embed}$ is the token embedder, $f_{ts}$ is the time series embedding layer, $L_{txt}$ is the length of prompt, $L_{ts}$ is the length of time series, and $D_{hidden}$ is the hidden dimension of embeddings.

Spectral-based GNNs such as Graph Convolutional Networks (GCN) cannot directly process directed acyclic graphs [36]. These models are inherently designed for undirected graphs. The only feasible approach to apply them to directed graphs is to symmetrize the adjacency matrix thereby converting the directed graph into an undirected approximation [37]. To encode the causal relationship graph represented as a directed acyclic graph obtained from the prior oriented PC algorithm into informative vector representations, we employ a Dense Graph Propagation (DGP) module [38].

In DGP, the adjacency matrix is decomposed into ancestor connections and descendant connections according to the direction of the edges. The propagation is then computed separately for each direction to capture the distinct influence flows. An asymmetric normalization scheme is further applied to preserve the original directionality of the edges, ensuring that the causal structure of the graph is accurately represented during the message passing process.

The process of causal graph embedding can be formally formulated as follows. Notably, to align with temporal dynamics, the time series are incorporated as node features and jointly fed into the graph embedding layer.

$$G_c = (\mathcal{V}, \mathcal{E}), \text{ where } |\mathcal{V}| = F \quad (9)$$

$$E_{causal} = f_{gnn}(\tilde{\mathbf{X}}, G_c) \in \mathbb{R}^{B \times L_{ts} \times D_{causal}} \quad (10)$$

$$E'_{causal} = f_{graph}(E_{causal}) \in \mathbb{R}^{B \times L_{ts} \times D_{hidden}} \quad (11)$$

where $G_c$ is the causal graph, $\mathcal{V}$ is the node set, $\mathcal{E}$ is the edge set, $F$ is the number of variables, $f_{gnn}$ is the DGP layer, and $f_{graph}$ is linear layer.

## D. Multimodal Information Fusion

To effectively integrate the modalities of time series data, task prompts, and causal graphs, we fuse the three distinct embeddings obtained through the data adaptation process into a unified feature vector that serves as the comprehensive input to the LLM. Specifically, we combine the time series embedding and the causal embedding via element-wise addition. This design is motivated by the observation that causal embeddings can be naturally interpreted as high-level features derived from the temporal dynamics of the data. The resulting fused embedding thus captures not only the intrinsic temporal evolution patterns within the time series but also the causal relationships among the variables.

In contrast to the causal embedding, which encodes intrinsic data characteristics, the prompt embedding represents textual descriptions regarding the task type and specific instructions. An effective prompt embedding enriches the input context, enabling the LLM to better comprehend the intended task. To this end, we adopt a "Prompt-as-Prefix" strategy, in which the prompt embedding is concatenated with the time series embedding in a prefix fashion [3]. Given that our tokenizer is kept frozen during training, this approach ensures seamless compatibility and interpretability of the prompt embeddings by the LLM. The fusion process can be summarized as:

$$E_{fused} = E_{ts} + E'_{causal} \in \mathbb{R}^{B \times L_{ts} \times D_{hidden}} \quad (12)$$

$$E_{input} = \text{Concat}(E_{txt}, E_{fused}) \in \mathbb{R}^{B \times (L_{txt} + L_{ts}) \times D_{hidden}} \quad (13)$$

## E. CM-LLM Inference

The multimodal fused embeddings $E_{input}$ are fed as inputs to the LLM, which generates initial hidden state sequence $H_{outpt}$ in an end-to-end process. The portion of the hidden state sequence corresponding to the time series is isolated by slicing, leveraging the known length of the instructional prompt $L_{txt}$. As shown in Fig. 3, the segmented hidden states $H_{ts}$ are passed through a projection layer, $f_{proj}$, which projects them back to the original feature space to produce the final prediction $\hat{\mathbf{Y}}$.

$$H_{output} = LLM(E_{input}) \in \mathbb{R}^{B \times (L_{txt} + L_{ts}) \times D_{hidden}} \quad (14)$$

$$H_{ts} = H_{output}[:, L_{txt} :, :] \in \mathbb{R}^{B \times L_{ts} \times D_{hidden}} \quad (15)$$

$$\hat{\mathbf{Y}} = f_{proj}(H_{ts}) \in \mathbb{R}^{B \times L_{ts} \times F} \quad (16)$$

## F. CM-LLM Fine-Tuning Process

### 1) Model Supervised Fine-Tuning

In SFT, LLM is trained with specially designed loss functions. The loss function includes 2 terms: the accuracy loss ($L_{acc}$) and the mask loss ($L_{mask}$), as shown in (17)–(18). $\lambda_1$ is the weight to balance. $L_{acc}$ is a comprehensive loss to understand the changing pattern of time series data. $L_{mask}$ employs mean squared error to minimize the point-to-point discrepancies in the mask segments.

$$\min\{L_{acc} + \lambda_1 L_{mask}\} \quad (17)$$

$$L_{acc} = \left\| \hat{\mathbf{Y}} - \mathbf{X} \right\|_2^2, L_{mask} = \left\| \hat{\mathbf{Y}} \circ \mathbf{M} - \mathbf{X} \circ \mathbf{M} \right\|_2^2 \quad (18)$$

To reduce the required training resources, we adopt a parameter-efficient fine-tuning approach known as Low Rank Adaptation (LoRA), which introduces a limited set of trainable parameters to adapt the pretrained model [39]. For a pretrained weight matrix $W_0 \in \mathbb{R}^{d \times k}$, its update is constrained through a low rank decomposition:

$$W = W_0 + T(W_0) = W_0 + BA \quad (19)$$

where $B \in \mathbb{R}^{d \times r}$ and $A \in \mathbb{R}^{r \times k}$, with the rank $r \ll \min(d, k)$. This low rank approximation effectively reduces the number of trainable parameters while preserving the expressive capacity of the model during adaptation.

### 2) The Dynamic Input-output Padding Mechanism

It is worth noting that in different task settings, the lengths of input and output sequences can vary significantly. To effectively address the inherent variability in sequence lengths across multiple tasks in time series learning, our framework adopts a hierarchical strategy for both sequence processing and loss computation, as shown in Fig. 6. During the data preprocessing stage, all input sequences are standardized to a predefined fixed length $L_{fix}$ through padding or truncation. This ensures structural consistency in the input space for subsequent deep learning components.

In contrast, the output sequence lengths may differ across tasks, while the output dimension of the projection layer remains fixed. To reconcile this discrepancy, we employ a padding-and-slicing mechanism. Specifically, during label loading, a custom collate function dynamically pads the output labels within each batch to match the maximum sequence length $L_{max}^{batch}$ present in that batch. This enables efficient batched computation. At the same time, the original valid sequence length $L_{valid}$ for each sample is preserved as metadata and passed along with the batched data into the training pipeline. During the loss computation phase, we implement a sequence-aware approach: although the model output and padded labels may not align in length, the loss function uses $L_{valid}$ to slice both predictions and ground truths, computing the error only over the valid portion of each sequence.

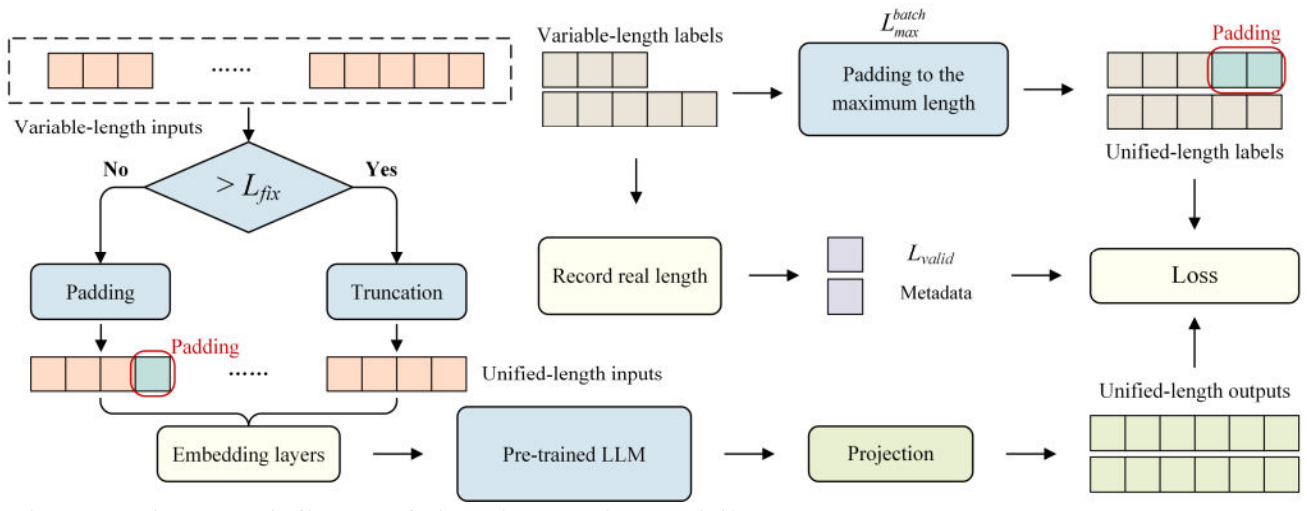

Fig. 6. The workflow of the dynamic padding process.

This strategy precisely isolates the influence of padding values on gradient backpropagation. As a result, our method enables efficient batch processing of variable-length sequences while overcoming the limitations imposed by the rigid structure of the model architecture.

## IV. CASE STUDY

### A. Experiment Settings

In this section, we conduct a comprehensive evaluation of the proposed LLMs across various tasks. The Qwen model is selected as the backbone pre-trained model due to its well-documented stability and broad applicability. To investigate the impact of model scale, we evaluate variants with approximately 0.6B, 1.7B, 4B, and 8B parameters. A series of ablation experiments are further carried out to analyze the contribution of different components and to better understand the model behavior. All models are fine-tuned for 50 epochs on NVIDIA A100 GPUs under a consistent training setup.

The training dataset used in this case study consists of real-world operational data with a temporal resolution of one minute, collected from a photovoltaic (PV) power plant equipped with an energy storage system. A total of 365 daily records are selected to construct the SFT training dataset, which includes six key features: solar irradiance, ambient temperature, PV output power, total output power, energy storage output, and the battery's state of charge (SOC).

Solar irradiance and temperature have a direct impact on the power generation of the PV system. The PV and energy storage systems operate in a coordinated manner, where the PV unit generates electricity unidirectionally based on available sunlight, while the energy storage system enables bidirectional regulation. When PV output exceeds the load demand, the excess energy is stored by charging the battery, resulting in SOC increase. Conversely, when PV output is insufficient to meet the demand, the energy storage system actively discharges to compensate for the deficit, leading to a decrease in SOC. Based on prior domain knowledge, this causal relationship among the variables is abstracted and represented as the graphical structure illustrated in Fig. 7.

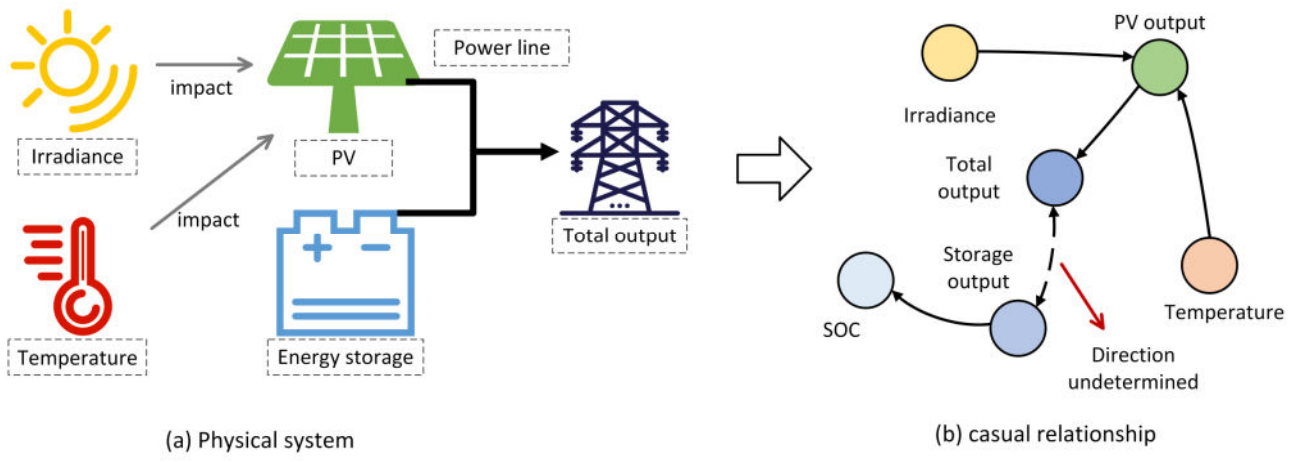

Fig. 7. (a) The influence relationship of the physical photovoltaic power plant. (b) The physics-statistics combined causal relationship derived from physical systems. The causal direction between the total output and storage output is identified through the causal discovery.

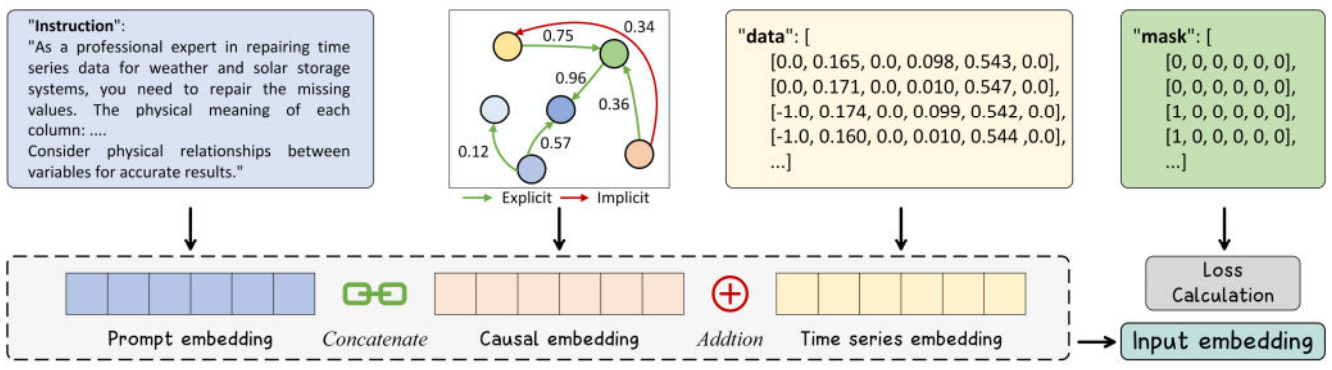

Fig. 8. A sample in the SFT dataset includes a prompt, a causal graph, a multivariate time series matrix and a mask matrix.

A concrete sample under the imputation scenario is illustrated in Fig. 8. The prompt, serving as a prefix, incorporates role-playing and task instructions to specify the task, data types, and other contextual information, thereby enhancing the LLM's adaptability to downstream tasks. The causal relationships are represented as a directed weighted graph, where nodes denote variables and edges indicate the strength and direction of causal influences. The time series data and corresponding mask are arranged in two-dimensional matrixes. After embedding, the time series form part of the model input. And the mask matrix is used for loss computation.

To evaluate the performance of the unified model across different tasks, we selected three specific task scenarios:

- **Imputation:** Many datasets exhibit consistent temporal granularity; for example, one-minute resolution yields 1,440 samples per day. We perform full sequence reconstruction but replace only missing values with their estimated counterparts from the model output. This preserves a fixed output length matching the original, avoiding complications from variable-length sequences due to missing data.
- **Forecasting:** Predictions are often needed across various time horizons, from ultra-short-term to long-term. Longer horizons pose greater challenges. We evaluate performance in two settings: hour-ahead and day-ahead forecasting.
- **Super-resolution:** Different downstream applications may require different levels of granularity. We define two super-resolution tasks: one that upscales the 15-minute interval data to 5-minute resolution, and another that enhances the 5-minute interval data to 1-minute resolution.

As comparisons, we train benchmarking models for every task: Multi-layer Perceptron (MLP), LSTM, Bidirectional LSTM (Bi-LSTM), Stacked Auto-Encoder (SAE), Transformer, and TimesNet [40]. These baselines are trained for 500 epochs.

### B. Performance Comparison with Baselines

CM-LLM's performance was evaluated using both point-wise metrics to assess reconstruction accuracy and distribution-wise metrics to quantify statistical fidelity, including Root Mean Square Error (RMSE) and Mean Absolute Error (MAE), Fréchet Inception Distance (FID) and Dynamic Time Warping (DTW).

TABLE II
RESULTS OF MODEL PERFORMANCE FOR DIFFERENT TASKS

| Task | Model | MLP | | LSTM | | Bi-LSTM | | SAE | | Transformer | | TimesNet | | Proposed (0.6B) | | Improvement | |
|---|---|---|---|---|---|---|---|---|---|---|---|---|---|---|---|---|---|
| | Metric | MAE | RMSE | MAE | RMSE | MAE | RMSE | MAE | RMSE | MAE | RMSE | MAE | RMSE | MAE | RMSE | MAE | RMSE |
| Imputation | all points | 0.1033 | 0.1583 | 0.0501 | 0.1518 | 0.0416 | 0.0805 | 0.0893 | 0.1518 | 0.0363 | 0.0799 | 0.0530 | 0.1295 | **0.0355** | **0.0731** | 2.20% | 8.51% |
| | masked | 0.1288 | 0.1912 | 0.1231 | 0.1847 | 0.0989 | 0.1579 | 0.1133 | 0.1847 | 0.0859 | 0.1453 | 0.1139 | 0.1637 | **0.0368** | **0.0748** | 57.40% | 48.59% |
| Forecasting | day-ahead | 0.0996 | 0.1893 | 0.1118 | 0.2189 | 0.1261 | 0.2259 | 0.0910 | 0.1708 | 0.1210 | 0.1780 | 0.0944 | 0.1660 | **0.0649** | **0.1045** | 31.24% | 37.05% |
| | hour-ahead | 0.0455 | 0.0970 | 0.0874 | 0.0951 | 0.0827 | 0.1230 | 0.0432 | 0.0951 | 0.0832 | 0.1251 | **0.0377** | **0.0864** | 0.0391 | 0.0886 | -0.51% | -1.49% |
| Super-resolution | 15min→5min | 0.0356 | 0.0508 | 0.0347 | 0.0591 | 0.0283 | 0.0535 | 0.0361 | 0.0525 | 0.0383 | 0.0561 | 0.0340 | 0.0578 | **0.0252** | **0.0470** | 10.95% | 12.15% |
| | 5min→1min | 0.0571 | 0.0733 | 0.0576 | 0.0838 | 0.0470 | 0.0713 | 0.0612 | 0.0779 | 0.0624 | 0.0818 | 0.0672 | 0.0799 | **0.0458** | **0.0685** | 2.55% | 3.93% |

*Bold term indicates the best performance, while underlining term represents second best.

TABLE III
FID AND DTW OF SUPER-RESOLUTION

| Model | MLP | | LSTM | | Bi-LSTM | | SAE | | Transformer | | TimesNet | | Proposed (0.6B) | |
|---|---|---|---|---|---|---|---|---|---|---|---|---|---|---|
| Metric | FID | DTW | FID | DTW | FID | DTW | FID | DTW | FID | DTW | FID | DTW | FID | DTW |
| Irradiance | 8.0628 | 1.6512 | 5.4604 | 1.6179 | 4.9163 | 1.3034 | 7.3341 | 1.8810 | 5.3947 | 1.5209 | 6.8100 | 1.3513 | **3.7903** | **0.8418** |
| Temperature | 1.5941 | 0.8430 | 1.2371 | 0.5584 | 0.6426 | 0.5479 | 1.5514 | 0.6325 | 0.7587 | 0.7966 | 1.3078 | 1.3749 | **0.5228** | **0.4022** |
| PV output | 4.4366 | 1.7778 | 3.8346 | 1.3936 | 2.9828 | 1.3476 | 4.0297 | 1.6184 | 2.9970 | 1.3651 | 4.2114 | 0.1775 | **2.8796** | **0.9314** |
| Total output | 8.8197 | 1.4567 | 7.9281 | 1.3152 | 4.7414 | 1.1995 | 7.1753 | 1.5104 | 3.8391 | 1.7234 | 6.2365 | 1.5118 | **3.2489** | **1.1447** |
| Storage output | 6.3652 | 1.3679 | 6.2505 | 1.3042 | 4.4824 | **1.1902** | 6.5175 | 1.3700 | 4.0381 | 1.6164 | 5.9920 | 1.3205 | **3.7476** | 1.2217 |
| SOC | 4.5849 | 0.6782 | 2.9319 | 0.6861 | **1.8116** | 0.6557 | 3.9521 | 0.8559 | 2.4570 | 0.7394 | 2.6179 | 0.8983 | 2.1845 | **0.5128** |

*Bold term indicates the best performance, while underlining term represents second best.

$$RMSE = \sqrt{\frac{1}{n}\sum_{i=1}^{n}(y_i - \hat{y}_i)^2} \tag{20}$$

$$MAE = \frac{1}{n}\sum_{i=1}^{n}|y_i - \hat{y}_i| \tag{21}$$

$$\mathrm{FID} = \| \mu_{\mathrm{real}} - \mu_{\mathrm{gen}} \|_2^2 + \mathrm{Tr}\left(\Sigma_{\mathrm{real}} + \Sigma_{\mathrm{gen}} - 2(\Sigma_{\mathrm{real}}\Sigma_{\mathrm{gen}})^{1/2}\right) \tag{22}$$

$$DTW(R,G) = \min_{W} \sum_{s=1}^{k} d(w_s) \tag{23}$$

where $y_i$ is the ground truth, $\hat{y}_i$ is the output data and $n$ is the number of data points. In equation (22), $\mu$ is the mean of feature vectors, $\sum$ is the covariance matrix and $\mathrm{Tr}(\cdot)$ is the trace of a matrix. In equation (23), $d(w_s)$ is the distance between two points, and $w_s=(i, j)$ represents the correspondence between the $i$-th point of real data $R$ and the $j$-th point of generated data $G$.

As summarized in Table II and III, the CM-LLM demonstrates superior performance over established baselines across most tasks. The most significant gains are observed in imputation, attributed to the effective leveraging of causal analysis, and in day-ahead forecasting, confirming its reliable long-term predictive capability. While the model shows a negligible performance gap in simpler hour-ahead tasks and its super-resolution advantage diminishes at finer temporal resolutions, this is deemed an acceptable trade-off. The unified architecture inherently sacrifices some task-specific optimization to achieve broad applicability, demonstrating overall robustness in capturing complex temporal dependencies. In the super-resolution task, the method's superiority is validated by consistently lower FID and DTW values, indicating strong fidelity in reconstructing data distributions and preserving temporal dynamics. Notably, the proposed model maintains its performance advantage for highly stochastic variables (e.g., solar irradiance). This robust performance in challenging scenarios further validates the model's advanced capability.

## C. *Performance Comparison with Different Model Sizes*

Scaling laws suggest that larger language models tend to achieve better performance. To investigate this relationship within our framework, we conduct experiments using four model sizes from the Qwen3 series: 0.6B, 1.7B, 4B, and 8B parameters. The training results, shown in Fig. 9, provide insights into how model capacity influences performance under our setup.

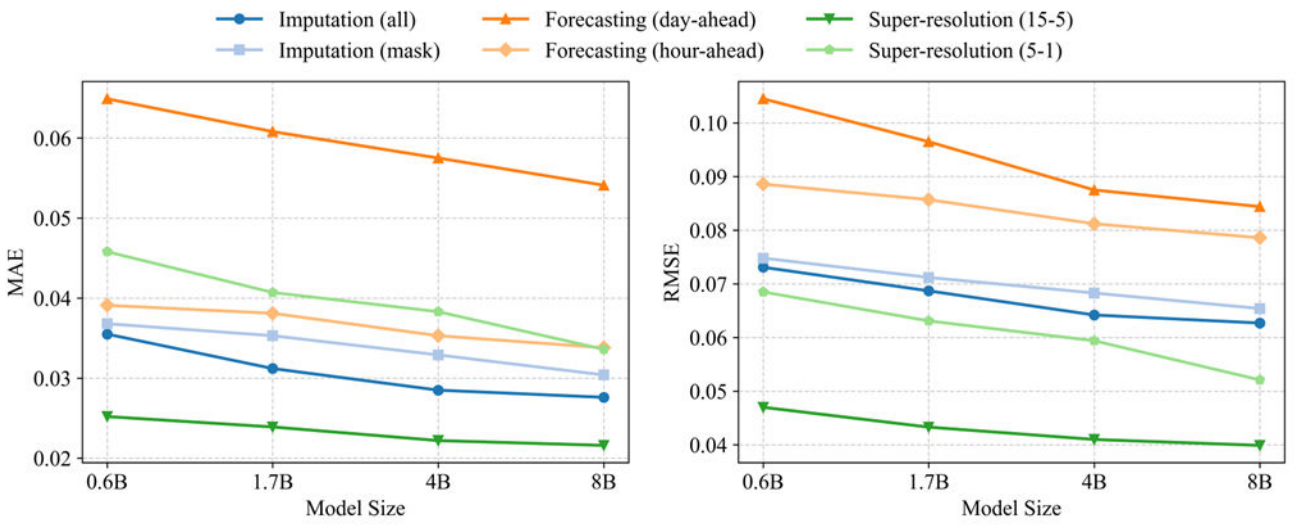

Fig. 9. The performances of different model sizes in tasks.

Both MAE and RMSE metrics exhibit a consistent downward trend as the model size increases. Specifically, for the MAE metric, the error decreases monotonically across all tasks as the model size grows from 0.6B to 8B parameters. For instance, the MAE for forecasting (day-ahead) drops from approximately 0.07 at 0.6B parameters to around 0.05 at 8B parameters. Similarly, the RMSE shows a parallel decrease, with the largest reduction observed in the forecasting (day-ahead) task, where the RMSE declines from about 0.10 at 0.6B parameters to roughly 0.08 at 8B parameters. This trend indicates that increasing the model capacity leads to improved performance across various tasks, which is consistent with the scaling laws hypothesis. The diminishing returns observed as the model size increases from 4B to 8B parameters suggest that while larger models continue to offer performance gains, the marginal improvements become smaller. This finding underscores the importance of balancing model size with computational resources and highlights the potential for further optimization in model architecture and training strategies to maximize performance gains with increasing model capacity.

## D. *Ablation Study*

To better understand the contribution of each model component, we conduct ablation studies by systematically removing specific elements and evaluating the resulting performance. In particular, we focus on the influence of prompt data and causal relationship graphs. For illustrative

TABLE IV
RESULTS OF ABLATION STUDY

| Task | Model | None | | | | Prompt-only | | | | Causal-only | | | | Proposed | | | |
|---|---|---|---|---|---|---|---|---|---|---|---|---|---|---|---|---|---|
| | Metric | MAE | RMSE | $D$ | $P_{MAE}$ | MAE | RMSE | $D$ | $P_{MAE}$ | MAE | RMSE | $D$ | $P_{MAE}$ | MAE | RMSE | $D$ | $P_{MAE}$ |
| Imputation | all points | 0.043 | 0.068 | 2.336 | 0.058 | 0.032 | 0.046 | 2.321 | 0.050 | 0.034 | 0.048 | **2.236** | 0.049 | **0.036** | **0.073** | 2.253 | **0.047** |
| | masked | 0.049 | 0.083 | | | 0.040 | 0.077 | | | 0.039 | **0.073** | | | **0.037** | 0.075 | | |
| Forecasting | day-ahead | 0.087 | 0.142 | 1.638 | 0.038 | 0.073 | 0.116 | 1.527 | 0.033 | 0.074 | 0.132 | 1.381 | 0.036 | **0.065** | **0.104** | **1.138** | **0.026** |
| | hour-ahead | 0.053 | 0.097 | 2.523 | 0.028 | **0.037** | **0.083** | 2.346 | 0.027 | 0.050 | 0.092 | 2.278 | **0.015** | 0.039 | 0.089 | **2.114** | 0.016 |
| Super-resolution | 15min→5min | 0.046 | 0.047 | 1.454 | 0.023 | 0.031 | 0.044 | 1.386 | 0.018 | 0.042 | 0.058 | 1.139 | 0.016 | **0.025** | **0.047** | **0.875** | **0.013** |
| | 5min→1min | 0.051 | 0.079 | 2.317 | 0.031 | 0.041 | 0.062 | 2.194 | 0.029 | 0.048 | 0.072 | 1.900 | **0.022** | **0.046** | **0.068** | **1.815** | 0.024 |

*Prompt-only: the proposed method without causal graphs. Causal-only: the proposed method without prompt data. Bold term indicates the best performance, while underlining term represents second best.

purposes and to maintain experimental clarity, all analyses are carried out using the Qwen-0.6B.

To evaluate model fidelity beyond point-wise accuracy, we assess its ability to preserve two critical structural properties of the multivariate time series data: statistical inter-dependencies among variables and inherent physical constraints of the system. To this end, we introduce two specialized metrics:

1. Correlation Discrepancy $D$: This metric quantifies the model's capacity to replicate the statistical dependencies among variables. It is calculated as the sum of the absolute differences between the elements of the Pearson correlation matrix derived from the model output and that from the ground truth. A lower value indicates a more accurate representation of the inter-variable relationships.

2. Power Balance Mean Absolute Error $P_{MAE}$: This metric assesses the model's adherence to the physical law of power conservation (i.e., *Output = PV + Storage*). It is computed as the MAE between the power balance residual series derived from the model's predictions and that from the ground truth.

$$D = \sum_{i=1}^{N-1} \sum_{j=i+1}^{N} \left| \underset{x \in \mathbf{X_o}}{Corr}(x_i, x_j) - \underset{x \in \mathrm{X_g}}{Corr}(x_i, x_j) \right| \tag{24}$$

$$P_{MAE} = \frac{1}{n} \sum_{i=1}^{n} |R_i^{true} - R_i^{pred}| \tag{25}$$

$$R_i = P_i^{total} - (P_i^{PV} + P_i^{Storage}) \tag{26}$$

where the $(x_i, x_j)$ is the feature pair, $\mathbf{X_o}$ is the output data, $\mathbf{X_g}$ is the ground truth, *Corr* means the correlation coefficient calculation function, $N$ is the number of feature and $n$ is the number of data points. The results summarize in Table IV.

The results demonstrate that the Causal Graph and prompt data modules have distinct, complementary functions. The Causal Graph module is instrumental in capturing the system's structural and physical properties, systematically reducing $D$ and $P_{MAE}$ across all tasks; for instance, in day-ahead forecasting, it lowered the D value from 1.527 to 1.138. Concurrently, the prompt data module primarily drives predictive accuracy, achieving substantial MAE and RMSE reductions compared to a causal-only variant, as evidenced by the 5min→1min super-resolution MAE drop from 0.029 to 0.022. This synergy validates the integrated architecture: the causal graph enforces physical plausibility via structural regularization, while the prompt data infuses rich, dynamic information to maximize precision, yielding a model that is simultaneously highly accurate and physically reliable.

### E. Cost Analysis

There is a trade-off between model parameter scale and training cost. As summarized in Table V, increases in parameter count consistently correlate with longer training durations, an effect attributed to greater operational complexity and memory consumption. A comparative analysis of the Prompt-only and Causal-only approaches reveals distinct scalability profiles. While the Prompt-only (tokenizer-based) method incurs lower overhead at smaller scales, the GNN-based Causal-only variant, though initially more demanding, demonstrates superior training efficiency at large scales (e.g., 4B and 8B). This suggests that the computational overhead of the GNN module remains relatively stable, whereas the cost associated with the tokenizer embedding scales substantially with model size. Consequently, the Causal-only architecture exhibits better overall scalability for large-scale implementations.

TABLE V
MODEL TRAINING COST

| Model | | Training time (h) | Training parameter | Training parameter percentage (%) |
|---|---|---|---|---|
| **0.6B** | None | 1.86 | 21248006 | 3.4362 |
| | Prompt-only | 2.64 | | |
| | Causal-only | 3.28 | 21314566 | 3.4462 |
| | Proposed | 4.58 | | |
| **1.7B** | None | 3.05 | 36983814 | 2.1017 |
| | Prompt-only | 4.16 | | |
| | Causal-only | 4.48 | 37116934 | 2.1090 |
| | Proposed | 5.56 | | |
| **4B** | None | 6.17 | 68706822 | 1.6783 |
| | Prompt-only | 8.89 | | |
| | Causal-only | 8.06 | 68873222 | 1.6822 |
| | Proposed | 9.72 | | |
| **8B** | None | 9.08 | 91524102 | 1.1045 |
| | Prompt-only | 11.81 | | |
| | Causal-only | 10.83 | 91790342 | 1.1076 |
| | Proposed | 12.67 | | |

## V. CONCLUSION

In this paper, we propose a causal-guided multimodal large language model (CM-LLM) to achieve generalized power system time series analytics. A physics-statistics combined causal discovery mechanism is proposed to capture the complex causal relationship among power system time series variables, and a multimodal data preprocessing framework is designed to achieve data fusion to enhance the model accuracy. Besides, a generic "mask-and-reconstruct" paradigm is proposed to solve the task specificity issue, while a dynamic input-output padding mechanism is designed to handle the structural rigidity problem. CM-LLM can fully unleash the generalization capabilities of LLMs in solving heterogeneous power system time series analysis tasks with superior accuracy and efficiency.

Three different time series tasks, i.e., missing data imputation, forecasting and super resolution are selected as examples to evaluate the model performances. Experimental results demonstrate that even when using Qwen3-0.6B as the backbone model, CM-LLM significantly outperforms conventional baseline approaches. Moreover, as the scale of the pre-trained LLM increases, the performance improves accordingly. We further conduct ablation studies to assess the contributions of task prompts and causal graph information, both of which are shown to play a critical role in enhancing model performances. However, it should be noted that performance gains tend to diminish as the model size increases to a certain level, while the training costs continue to rise significantly. Such observations highlight the need for a balanced trade-off between model scalability and efficiency.